\documentclass[11pt,twoside]{article}
\usepackage{asp2006}
\usepackage{psfig}
\usepackage{epsf}
\usepackage{graphics}
\usepackage{lscape}
\def\edcomment#1{\iffalse\marginpar{\raggedright\sl#1\/}\else\relax\fi}
\reversemarginpar

\begin{document}
\title{The Statistical Properties Of The Very Weak Radio Source Population In The GOODS/ACS HDF-N Region }
\author{T.W.B. Muxlow}
\affil{MERLIN-VLBI National Facility, Jodrell Bank Observatory, University of Manchester, United Kingdom}
\author{R.J. Beswick, H. Thrall, A.M.S Richards, S.T. Garrington, A. Pedlar}
\affil{Jodrell Bank Observatory, University of Manchester, United Kingdom}

\begin{abstract}
Deep combination radio observations at 1.4GHz with the VLA and MERLIN
have imaged a region 10 arcminutes square surrounding the Hubble Deep
Field North (HDF-N). Initial studies of the weak radio source
population have shown that the proportion of starburst systems
increases with decreasing radio flux density with more than 70\% of
radio sources being starburst in nature at flux densities less than
S${_{1.4{\rm GHz}}}$$\sim$70$\mu$Jy. The recently published GOODS ACS
field overlaps this area, and here we present the results of
a follow-up statistical study of the very weak radio sources
(S${_{1.4\rm GHz}}$$<$40$\mu$Jy) in an 8.5 arcminute square field
centred on the HDF-N which contains the region of overlap. Radio
emission at the level of a few $\mu$Jy are statistically detected
associated with ACS galaxies brighter than a z-band magnitude of
25. These very faint radio sources are extended starburst systems with
average radii in the range 0.6 to 0.8 arcseconds and for those with
measured redshifts, radio luminosities typically several times that of the
nearby well-studies starburst galaxy M82.

\end{abstract}

\vspace{-0.5cm}
\section{Introduction}

Deep galaxy studies at visible and infra-red wavelengths have
indicated that early galaxies merge to form larger systems in a
'bottom-up' scenario of galaxy assembly. This implies that
galaxy-galaxy interactions were common at early epochs. Such
interactions at modest redshifts are seen to trigger major
star-formation activity and it is well established that star-formation
rate density increases dramatically out to redshifts of $\sim$1
\citep{madau98}. The star-formation rate for such galaxies can be
estimated from both the centimetric radio and far infra-red (FIR)
luminosities which have been found to be tightly correlated over
several orders of magnitude
\citep{vanderkruit73,condon82}. 

Differential radio source counts at 1.4GHz (See Figure 1) show an
increase at flux densities below 1mJy
\citep{seymour04}. Multi-wavelength studies of a number of fields
including deep radio observations involving the VLA, ATCA, MERLIN, and
the EVN, have shown that this new weak radio source population is
associated with distant star-forming galaxies.

 \setcounter{figure}{0}

 \begin{figure}[!h]
 \plotfiddle{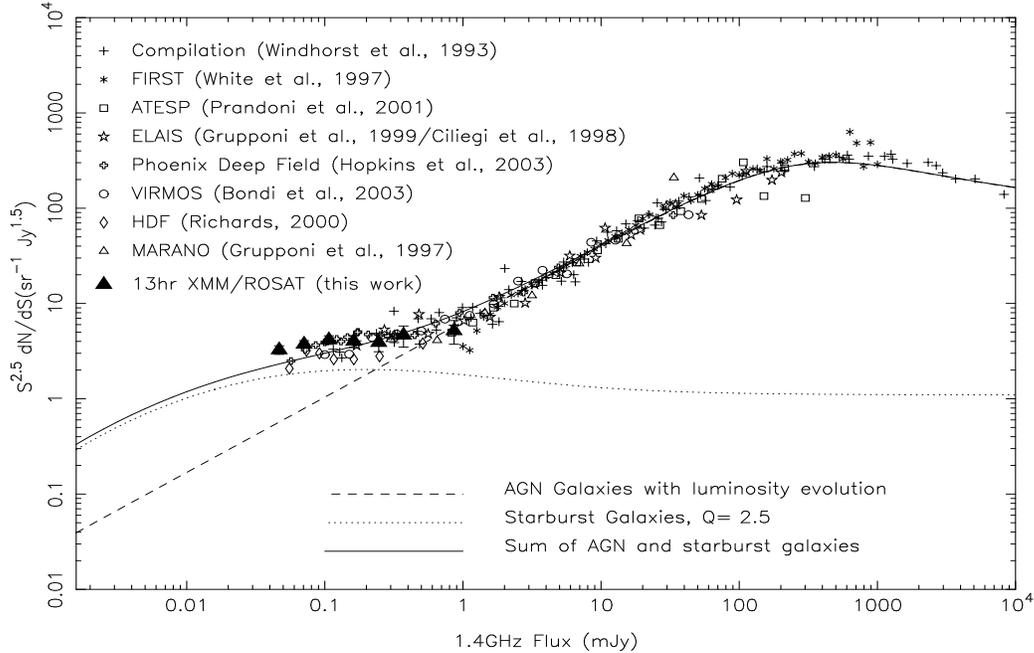}{7.5cm}{-90}{55}{48}{-210}{270}
\caption{Normalised differential radio source counts at 1.4GHz. At flux densities below 1mJy there is an upturn which is identified with a new population of distant star-forming galaxies. After Seymour, McHardy \& Gunn (2004).}
 \end{figure}

\section{The Initial Radio Study of the HDF-N}

Deep radio observations of the HDF-N region were made with the VLA and
MERLIN during 1996 and 1997. The VLA results were presented in
Richards ${\it et al.}$ (1998) and Richards (2000), and the
combination high angular resolution results were presented in Muxlow
${\it et \hspace{1mm} al.}$ (2005). Within a 10 arcminute square field centred on
the HDF, 92 radio sources were detected by the VLA at 1.4GHz above a
completeness limit of 40$\mu$Jy/beam (5.3$\sigma$) using a 2 arcsecond
beam. Combination high-resolution MERLIN+VLA images involving 42 hours
of VLA observations and 18 days of MERLIN data were made for small
regions around each of the 92 detected sources. The rms noise level in
the combination images was $\sim$3.3$\mu$Jy per 0.2 arcsecond beam
making these images some of the most sensitive yet produced at 1.4GHz.

The full results of the study of the individual radio sources in the
10 arcminute field surrounding the HDF-N are given in Muxlow ${\it et \hspace{1mm} al.}$ (2005) and are summarized below:

\noindent ${\bullet}$ Within the 10x10 arcminute field there are 92 radio
sources with flux densities at 1.4GHz $>$40$\mu$Jy.

\noindent ${\bullet}$ The radio sources have angular sizes in the range 0.2
to 3 arcseconds.

\noindent ${\bullet}$ 85\% of the radio sources are associated with galaxies brighter than 25${^{\rm th}}$ magnitude in the v-band.

\noindent ${\bullet}$ The remaining 15\% are optically faint or extremely
red objects lying at high redshift, some of which are detected at sub-mm
wavebands.

\noindent ${\bullet}$ The radio sources can be characterised as AGN or
starburst according to their radio structures with starburst systems being
identified as those with steep radio spectral indices and radio structures
extended on (sub-)galactic scales overlaying the central region of the
optical galaxy.

\noindent ${\bullet}$ Below 60$\mu$Jy the radio sources are dominated by
starburst type systems (See Figure 2).

\noindent ${\bullet}$ The starburst systems typically have radio luminosities
significantly greater than that of nearby well-studied starbursts such as M82
or Arp220.

\noindent ${\bullet}$ Some high redshift starburst systems show evidence for embedded
AGN \citep{garrett01}. 


 \begin{figure}[!h]
 \plotfiddle{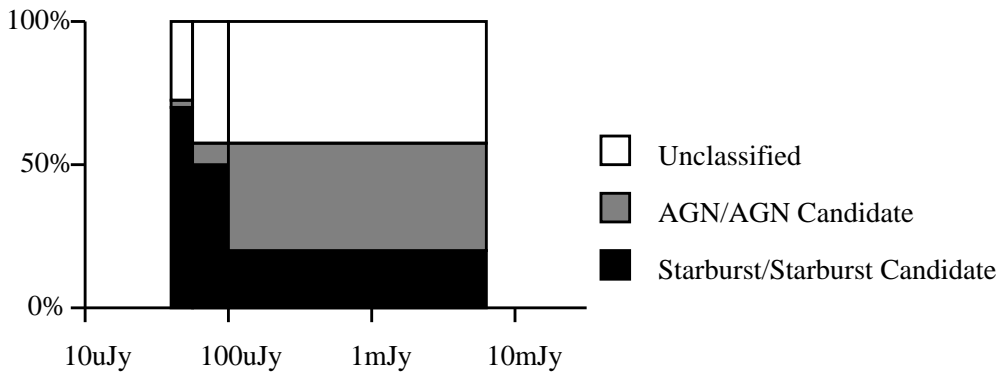}{4.5cm}{270}{60}{60}{-250}{250}
\caption{Distribution of source classification with flux density for the radio sources in the 10 arcminute field. After Muxlow {\it et al.}(2005).}
\end{figure}

\section{A New Study of the Very Weak Radio Source Population}

The initial investigation has been extended to include a statistical
study of the very weak radio source population in the HDF-N region
utilising ancillary data obtained from the large multi-wavelength
Great Observatories Origins Survey (GOODS). The GOODS North region
partially overlies the original field, and a new wide-field contiguous
combination radio image was made covering a complete 8.5 arcminute
square field centred on the HDF-N which contains the region of
overlap. The 8.5 arcminute square radio image covers a region of the
GOODS-N field containing 13030 galaxies brighter than magnitude 28.3
in the z-band ACS catalogue.

\subsection{Faint Radio Emission from Optically-selected Galaxies}

The radio and optical fields are astrometrically aligned to better
than 50mas. Thus the radio flux density from each of the 13030
z-band optical galaxies can be investigated by integrating the radio
emission from the MERLIN+VLA image of the 8.5 arcminute square field
within a radius of 0.75 arcseconds at the position of every
galaxy. Radio sources brighter than 40$\mu$Jy have already been
reported in detail in Muxlow ${\it et \hspace{1mm}  al.}$ (2005) and were excluded
from the sample. In addition, all galaxies with nearest neighbours
closer than 1.5 arcseconds were also excluded so as to avoid confusion
problems. The median flux densities, binned by z-band magnitude are
shown in Figure 3. A control sample was also constructed which
incorporated a random 7 arcseconds positional shift between the
optical galaxy and the position searched in the radio image.

 \begin{figure}[!h]
 \plotfiddle{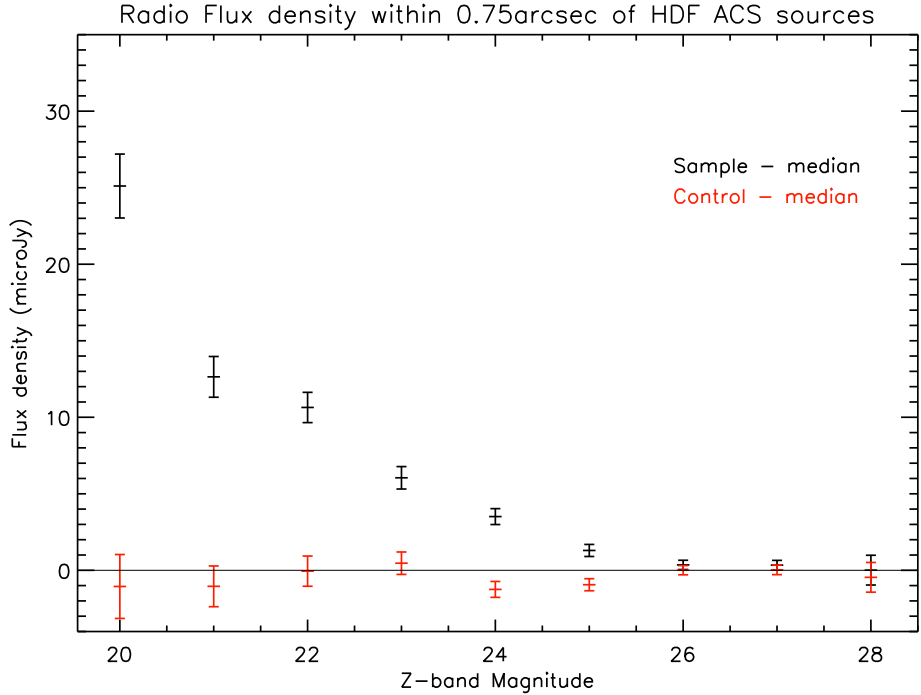}{9.0cm}{0}{75}{75}{-230}{-275}
\caption{The median 1.4GHz flux density detected within 0.75 arcseconds of the position of optical galaxies in the ACS HDF-N field, binned by z-band magnitude. The control sample incorporates a random 7 arcsecond shift.}
\end{figure}

Radio emission is clearly detected statistically for all z-band
magnitude bins brighter than 26. Of the $\sim$2700 galaxies brighter
than z=24${^{\rm mag}}$, half (nearly 1400) will have radio flux
densities of $\sim$4$\mu$Jy or greater at 1.4GHz. A 4$\mu$Jy source
represents an $\sim$8$\sigma$ detection for a future deep ${\it
e}$-MERLIN+EVLA image; such galaxies which at present may only be
investigated statistically will in the near future be imaged
individually.

\subsection{Radio Source Sizes}

The angular sizes of the very weak radio sources can also be
investigated statistically. In each magnitude bin the average radio
source size can be derived from the radio flux densities found in
annuli of width 0.25 arcseconds, over radii of 0.25 to 2 arcseconds
centred on each galaxy position. These are shown in Figure 4. For
detected systems brighter than z$\sim$25${^{\rm mag}}$, the average
radio source radii are in the range 0.6 to 0.8 arcseconds implying
that the next generation radio interferometers will need sub-arcsecond
angular resolution. There is a marginal trend for the fainter radio
sources to be slightly smaller. Radio source sizes may also be also be
measured by fitting to the composite, stacked radio image in each
magnitude bin. Similar sizes are derived by this latter method.

 \begin{figure}[!h]
 \plotfiddle{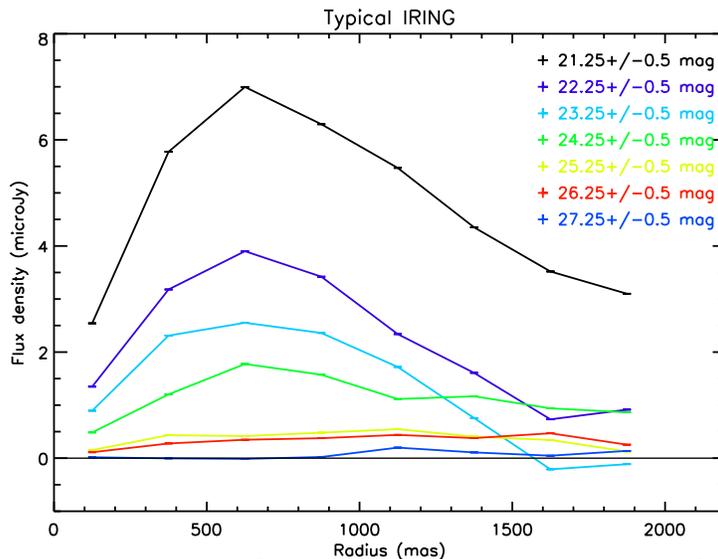}{6.0cm}{0}{60}{60}{-175}{-235}
\caption{The average radio flux density contained within annuli of width 0.25 arcseconds over radii of 0.25 to 2 arcseconds centred at the position on each galaxy binned by z-band magnitude. For galaxies brighter than z$\sim$25${^{\rm mag}}$, the average radio source radii are in the range 0.6 to 0.8 arcseconds.}
\end{figure}

\subsection{Starburst Luminosities}

Around 1000 of the $\sim$13000 galaxies in the 8.5 arcminute field
have published spectroscopic redshifts. We have constructed the
luminosity-redshift distribution for those starburst systems with
measured redshifts. For this sub-set of starbursts, the detected radio
flux density stacked by z-band magnitude does not differ significantly
from the complete galaxy sample (at least for those bins brighter than
24${^{\rm mag}}$); indicating that the sub-set of galaxies with measured
redshifts do not appear to differ significantly from the complete
galaxy set. We are thus able to derive the averaged luminosity
distribution for those starburst galaxies brighter than 24${^{\rm mag}}$

\begin{figure}
 \plotfiddle{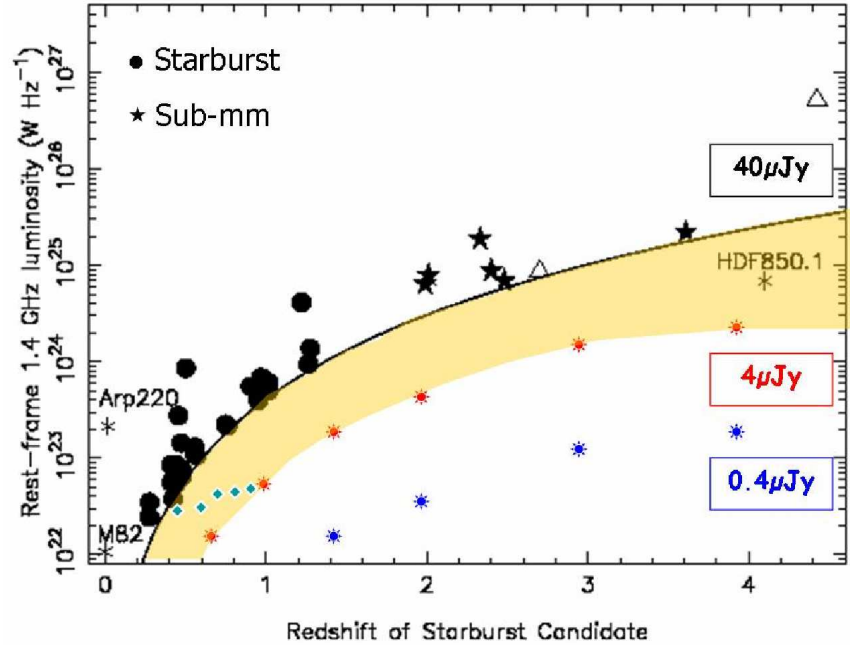}{7.5cm}{0}{35}{35}{-160}{-15}
\caption{Rest frame monochromatic luminosities for individual galaxies from Muxlow ${\it et \hspace{1mm} al.}$ (2005) and the statistical results presented here (filled diamonds). Sample cut-off levels of 40, 4, and 0.4$\mu$Jy are shown. For comparison, two nearby starbursts (M82 \& Arp220) are also shown together with HDF850.1, the brightest sub-mm ${\it SCUBA}$ detection in the HDF-N}
\end{figure}

The very weak radio sources which are assumed to be dominated by
starbursts have average properties which indicate that they are an
extension of the weaker end of the distribution of starbursts detected
as individual sources at higher radio flux densities. Their typical
luminosities are still several times that of M82. However, it should
be noted that these weak radio sources have been selected by optical
z-band properties and will therefore exclude optically faint
systems. Most of the individual high redshift sub-mm starburst systems
imaged by Muxlow ${\it et \hspace{1mm} al.}$ (2005) were optically faint, and
therefore will not be represented statistically in these
results. Furthermore close galaxy pairs have also been
excluded. Clearly individual imaging of these very faint systems will
be required. 

\section{The Future}

The new upgraded radio interferometers due to be commissioned in the
next few years will be able to individually image these very faint
systems. ${\it e}$-MERLIN, EVLA, and ${\it e}$-VLBI should be able to
image in excess of 1000 individual starburst systems to flux densities
of $\sim$4$\mu$Jy at 1.4GHz with perhaps 150 to 200 objects at high
redshift in a single field. For the even weaker radio source
population, the new instruments will be able to extend this present
statistical study to include many thousands of starburst systems with
radio flux densities less then 1$\mu$Jy. SKA and ALMA will ultimately
extend this by an additional order of magnitude, With redshifts,
improved spectral energy distribution templates and extinction-free
star-formation rate indicators, it will become possible to solve for
the cosmic star-formation history - the radio Madau diagram.





\begin{thebibliography}{}

\bibitem[\protect\citeauthoryear{Condon {\it et al.}}{1982}]{condon82} Condon, J. J., Condon, M. A., Gisler, G. \& Puschell, J., 1982, ApJ, 252, 102

\bibitem[\protect\citeauthoryear{Garrett {\it et al.}}{2001}]{garrett01} Garrett, M. A., {\it et al.}, 2001, A\&A, 366, L5

\bibitem[\protect\citeauthoryear{Madau, Pozzetti, \& Dickinson}{1998}]{madau98}
Madau, P., Pozzetti, L., \& Dickinson, M., 1998, ApJ, 498, 106

\bibitem[\protect\citeauthoryear{Muxlow {\it et al.}}{2005}]{muxlow05}
Muxlow, T. W. B., {\it et al.}, 2005 MNRAS, 358,
1159

\bibitem[\protect\citeauthoryear{Richards {\it et
al.}}{1998}]{richards98} Richards, E. A., Kellermann, K. I., Fomalont,
E. B., Windhorst, R. A., Partridge, R. B., 1998, AJ, 116, 1039

\bibitem[\protect\citeauthoryear{Richards\,}{2000}]{richards00}
Richards, E. A., 2000, ApJ, 533, 611

\bibitem[\protect\citeauthoryear{Seymour, McHardy, \& Gunn}{2004}]{seymour04} Seymour, N., McHardy, I.M., \& Gunn, K.F., 2004, MNRAS, 352, 131


\bibitem[\protect\citeauthoryear{van der Kruit\,}{1973}]{vanderkruit73} van der Kruit, P. C., 1973, A\&A, 29, 263





\end{thebibliography}
\end{document}